\documentclass[runningheads,natbib=false]{llncs}

\usepackage{graphicx}
\usepackage[ruled,vlined]{algorithm2e}
\usepackage[backend=bibtex,sorting=none]{biblatex}
\usepackage{xcolor}

\usepackage[normalem]{ulem} 

\usepackage{hyperref}
\usepackage{cleveref}

\def \superAgg {{PriceAggregator}}
\def \pb {{P_B(s_i)}}
\def \pa {{P_A(s_i)}}
\def \pd {{P_D(s_i)}}

\def \xgboost {{XGBoost}}
\def \agoda {{Agoda}}
\def \C {{\mathcal{C}}}
\def \smartTTL {{SmartTTL }}

\begin{document}
\title{PriceAggregator: An Intelligent System for Hotel Price Fetching}
%
%

\author{Jiangwei Zhang \and
Li Zhang \and
Vigneshwaran  Raveendran
\and 
Ziv Ben-Zuk \and 
Leonard Lu
}

\authorrunning{Jiangwei et al.}

\institute{Agoda Inc. \\
\email{jiangwei.zhang,li.zhang,vigneshwaran.raveendran,ziv.benzuk,Leonard.Lu@agoda.com}
}
\maketitle              
\begin{abstract}
This paper describes the hotel price aggregation system - \superAgg, 
deployed at \agoda, 
a global online travel agency for hotels, vacation rentals, 
flights and airport transfer. 
\agoda\ aggregates non-direct suppliers' hotel rooms to ensure that 
\agoda's customers always have the widest selection of hotels, room types and packages. 
As of today, \agoda\ aggregates millions of hotels.

The major challenge is that each supplier only allows \agoda\ to 
fetch for the hotel price with a limited amount of Queries Per Second (QPS). 
Due to the sheer volume of \agoda's user search traffic, 
this limited amount of QPS is never enough to cover the all user searches. 
Inevitably, many user searches have to be ignored. Hence, booking lost.

To overcome the challenge, we built \superAgg.
\superAgg\ intelligently determines when, 
how and what to send to the suppliers to fetch for price. 
In this paper, we not only prove \superAgg\ is optimal theoretically, 
but also demonstrate that \superAgg\ performs well in practice. 
\superAgg\ has been deployed in \agoda. 
Extensive online A/B experimentation 
have shown that \superAgg\ increases \agoda's bookings significantly.

\end{abstract}
\section{ Introduction }
\agoda~\footnote{\agoda.com} is a global online travel agency for 
hotels, vacation rentals, flights and airport transfers. 
Millions of guests find their accommodations and 
millions of accommodation providers list their properties in \agoda.
Among these millions of properties listed in \agoda, 
many of their prices are fetched through third party suppliers.

\begin{figure}[t]
    \centering
    \includegraphics[width = 4.3in]{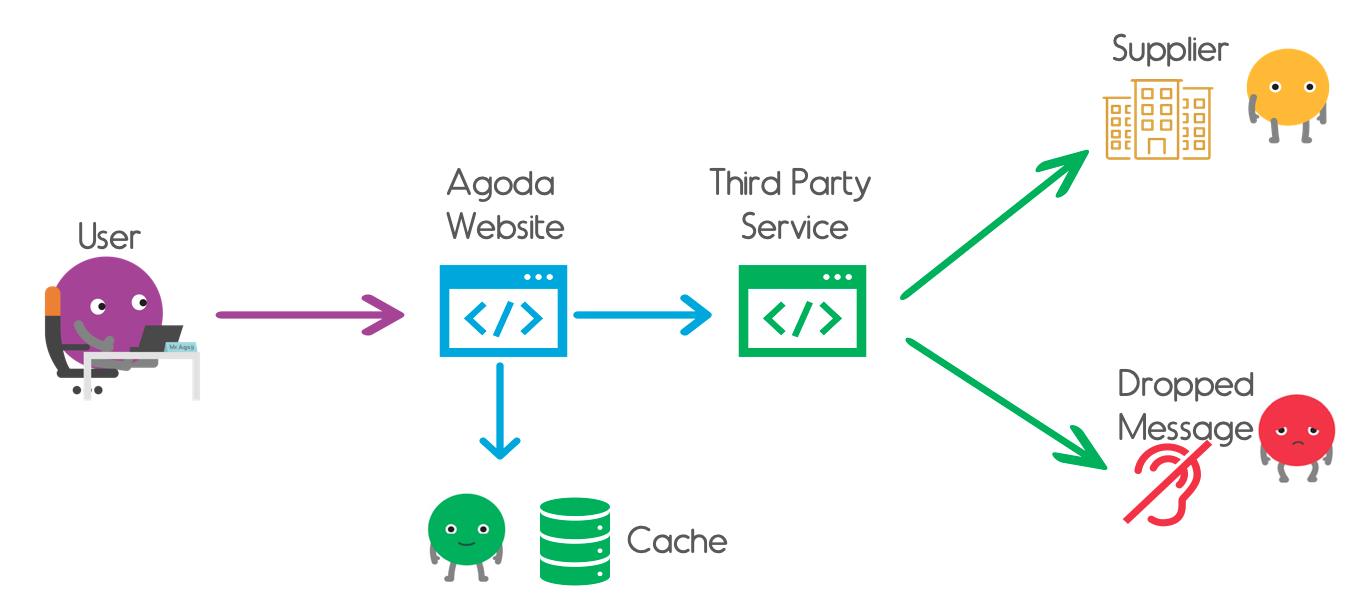}
    \caption{System flow of third party supplier hotel serving. 
    If a cached price exists, \agoda\ first serves the cache price to the user. 
    Otherwise, \agoda, on a best-efforts basis, 
    sends a request to the supplier to fetch for hotel price and put it in cache.}
    \label{fig:flow2}
\end{figure}

These third party suppliers 
do not synchronize the hotel prices to \agoda. 
Every time, to get a hotel price from these suppliers, 
\agoda\ needs to make 1 HTTP request call to the supplier to 
fetch the corresponding hotel price.
However, due to the sheer volume of the search requests received from users, 
it is impossible to forward every request to the supplier.
Hence, a cache database which temporarily stores the hotel prices is built. 
For each hotel price received from the supplier, 
\agoda\ stores it into this cache database for some amount of time and evicts the price from the cache once it expires. 
Figure~\ref{fig:flow2} above abstracts the system flow.

Every time an user searches a hotel in \agoda, 
\agoda\ first reads from the cache. 
If there is a hotel price for this search from the user in the cache, 
it is a 'hit' and we will serve the user with the cached price. 
Otherwise, it is a 'miss' and the user will not have the price for that hotel.
For every 'miss', 
\agoda\ will send a request to the supplier to get the price for that hotel, 
and put the returned price into the cache.
So that, the subsequent users can benefit from the cache price. 
However, every supplier limits the amount of requests we can send at every second. Once we reach the limit, the subsequent messages will be ignored. Hence, this poses four challenges.

\textbf{Challenge 1: Time-to-live (TTL) determination}.

For a hotel price fetched from the supplier, 
how long should we put such hotel price in the cache before expiring them? 
We call this duration as time-to-live (TTL). 
The larger the TTL, the longer the hotel prices stay in the cache database.
As presented in Figure~\ref{fig:ttl_role}, the TTL plays three roles:

\begin{figure}[t!]
    \centering
    \includegraphics[height=1.9in]{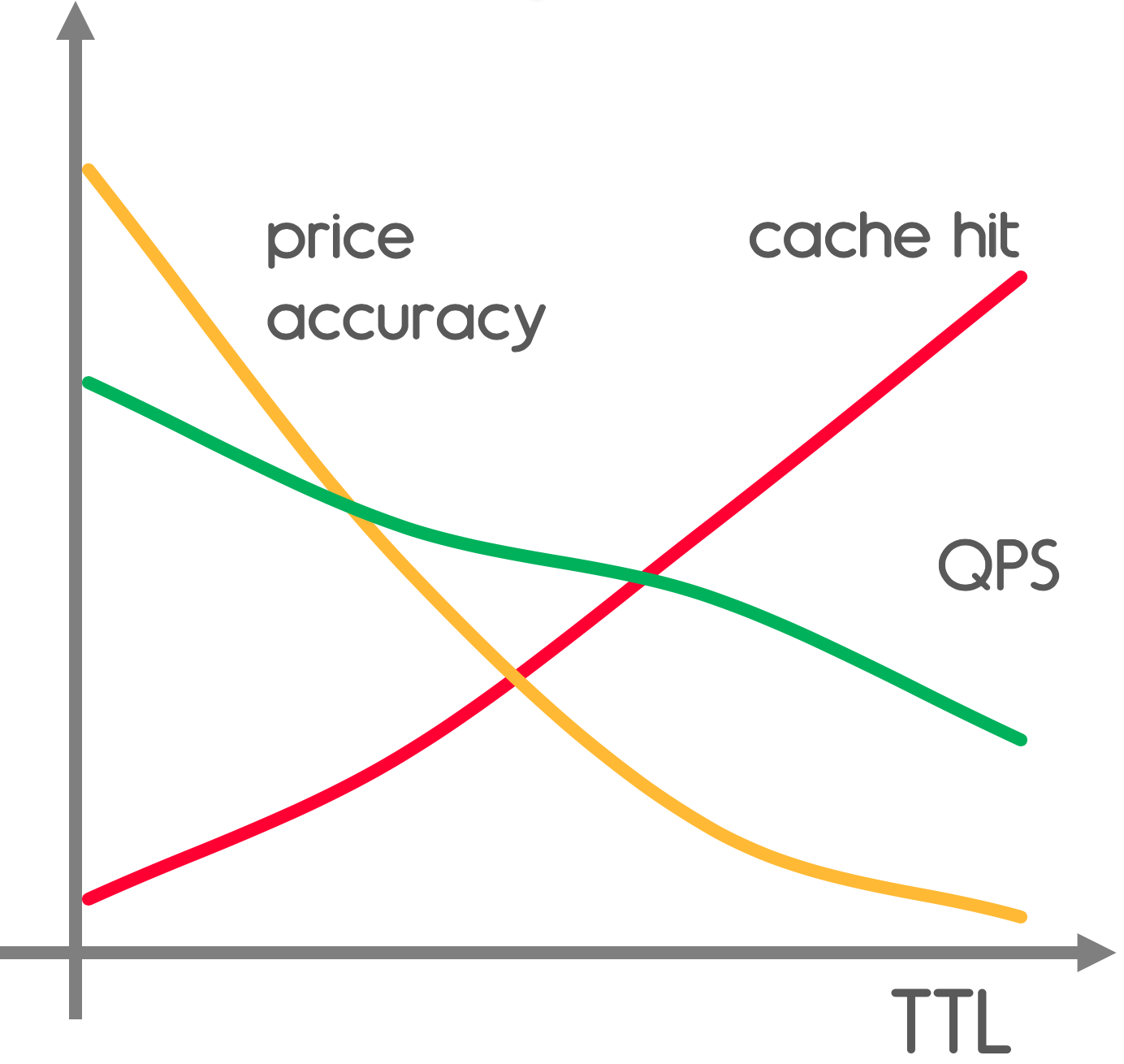}
    \caption{TTL v.s. cache hit, QPS and price accuracy.}
    \label{fig:ttl_role}
\end{figure}

\begin{itemize}
    \item \textbf{Cache Hit}.  
          With a larger TTL, hotel prices are cached in the database for a longer period of time and hence, more hotel prices will remain in the database. When we receive a search from our users, there is a higher chance of getting a hit in the database. This enhances our ability to serve our users with more hotel prices from the third party suppliers.
    
    \item \textbf{QPS}.
          As we have limited QPS to each supplier, a larger TTL allows more hotel prices to be cached in database. Instead of spending QPS on repeated queries, we can better utilise the QPS to serve a wider range of user requests.
    
    \item \textbf{Price Accuracy}.
  As the hotel prices from suppliers changes from time to time, a larger TTL means that the hotel prices in our cache database are more likely to be inaccurate. Hence, we will not be able to serve the users with the most updated hotel price.
          
\end{itemize}

There is a trade-off between cache hit and price accuracy. We need to choose the  TTL that caters to both cache hit and price accuracy.
To our best knowledge, 
most Online Travel Agents (OTA) typically pick 
a small TTL ranging from 15 minutes to 30 minutes. 
However, this is not optimal.

\textbf{Challenge 2: Cross data centre QPS management}.
 
    \begin{figure}
         \centering
    \includegraphics[height = 1.9in]{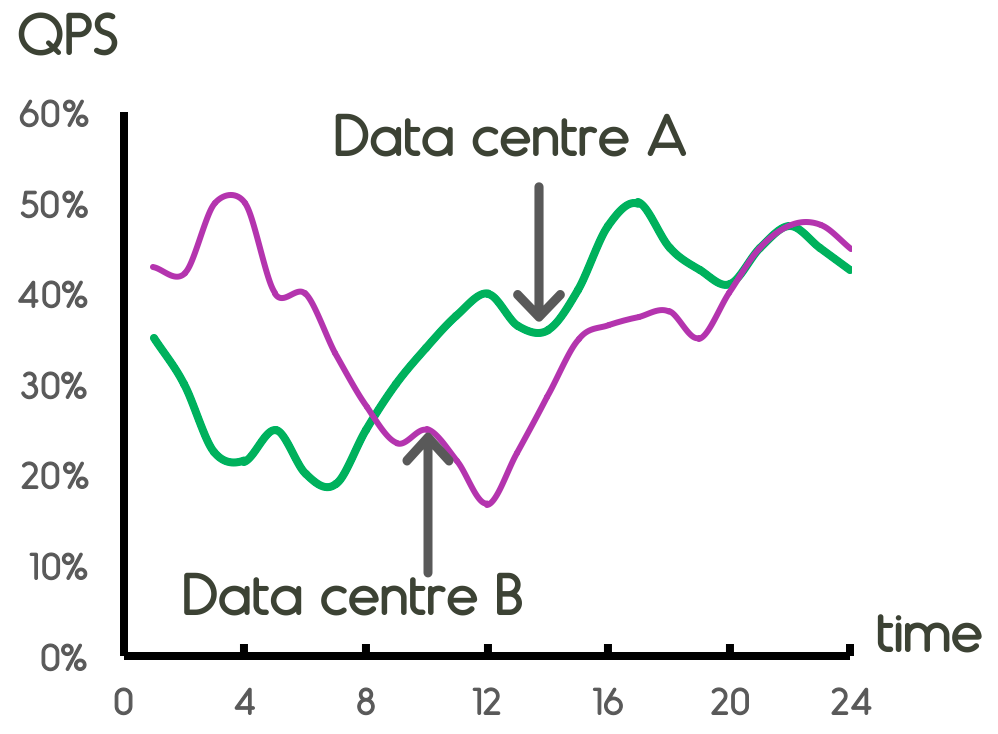}
    \caption{Cross data centre QPS management limitation. 
    Data centre A peaks around $50\%$ QPS around \texttt{18:00} and 
    data centre B peaks around $50\%$ QPS around \texttt{04:00}. }
    \label{fig:qps-multi-dc}
\end{figure}
\agoda\ has several data centres globally to handle the user requests. 
For each supplier,
we need to set a maximum number of QPS that each data centre is allowed to send.
However, each data centre has its own traffic pattern. 

Figure~\ref{fig:qps-multi-dc} presents an example of 
the number of QPS sent to a supplier from 
two data centres A and B.
For data centre A, it peaks around $50\%$ QPS around \texttt{18:00}.
At the same time, data centre B peaks around $50\%$ QPS around \texttt{04:00}.
If we evenly distribute this $100\%$ QPS to data centre A and data centre B, 
then we are not fully utilizing this $100\%$ QPS.
If we allocate more than $50\%$ QPS to each data center, 
how can we make sure that 
data center A and data center B never exceed the $100\%$ QPS in total?
Note that, the impact of breaching the QPS limit could be catastrophic to the supplier, 
which might potentially bring down the supplier to be offline.

\textbf{Challenge 3: Single data centre QPS utilization}.
    \begin{figure}
     \centering
    \includegraphics[height = 1.9in]{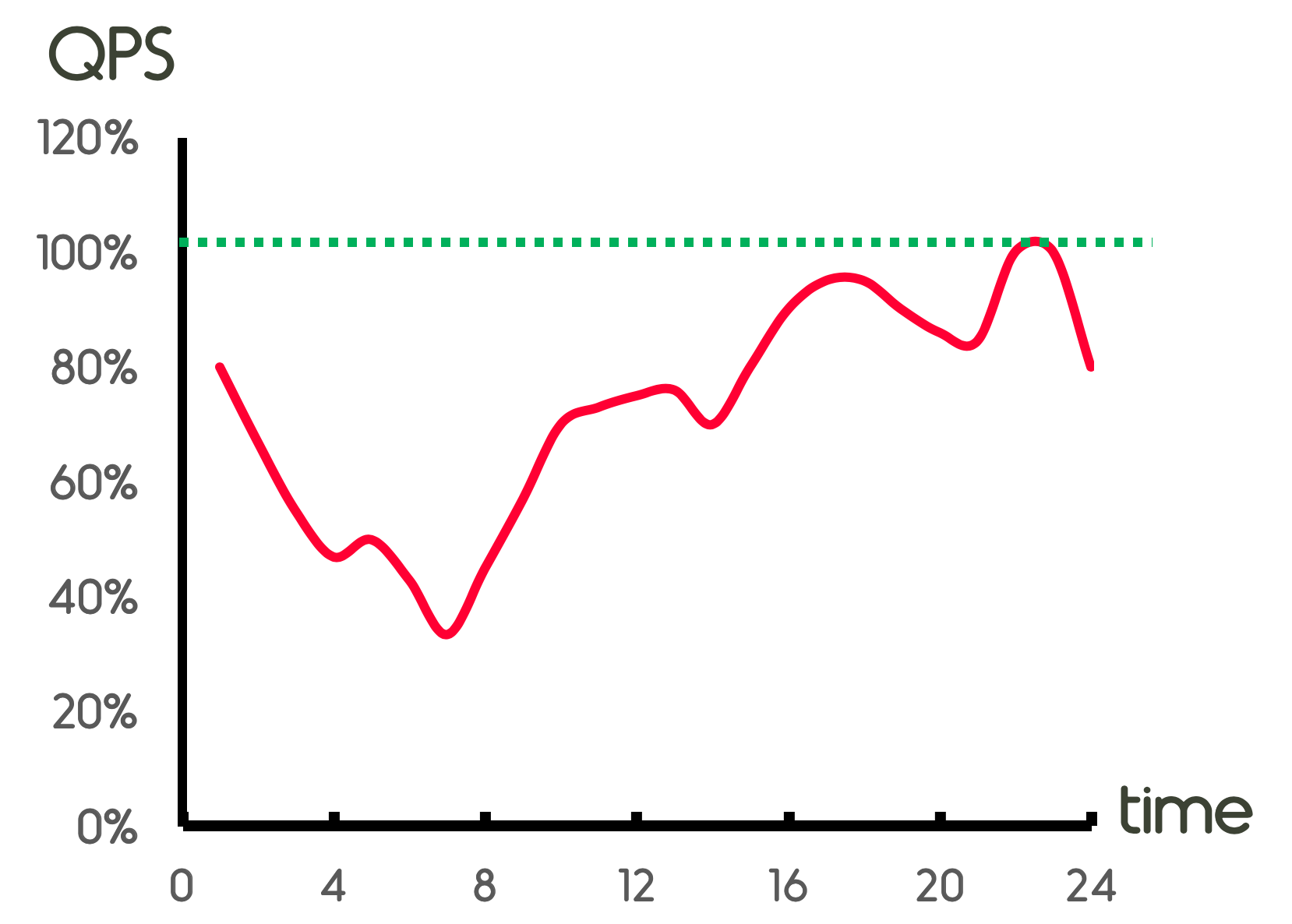}
    \caption{Un-utilized QPS}
    \label{fig:qps-utilization}
    \end{figure}
    
As mentioned in the previous section,
each data centre has its own traffic pattern, 
there are peak periods when we send the most amount of requests to the supplier, 
and non-peak period when we send much fewer number of requests to the supplier.
As demonstrated in Figure~\ref{fig:qps-utilization}, 
for this data centre, 
it sends \textless $40\%$ QPS to the supplier around \texttt{08:00}.
However, similar to the abovementioned example, $100\% - 40\% = 60\%$ QPS of this data centre is not utilized.

\textbf{Challenge 4: Cache hit ceiling}.

The passive system flow presented in Figure~\ref{fig:flow2} 
has an intrinsic limitation to improve the cache hit. 
Note that, this design sends a request to supplier to fetch for price
only if there is a miss. 
This is passive!
Hence, a cache hit only happens if 
the same hotel search happened previously and the TTL is larger than the 
time difference between the current and previous hotel search. 

Note that we cannot set TTL to be arbitrarily large as this will lower the price accuracy as explained in Challenge 1. As long as TTL of a specific search is not arbitrarily large, it will expire and the next request of this search will be a miss.
Even though we can set the TTL to arbitrarily large, 
those hotel searches that never happened before will always be miss. 
For example, if more than 20\% of the requests are new hotel searches. 
Then, it is inevitable for us to have \textless 80\% cache hit regardless of how large the TTL is set.

\begin{figure}[t]
    \centering
    \includegraphics[height=1.5in]{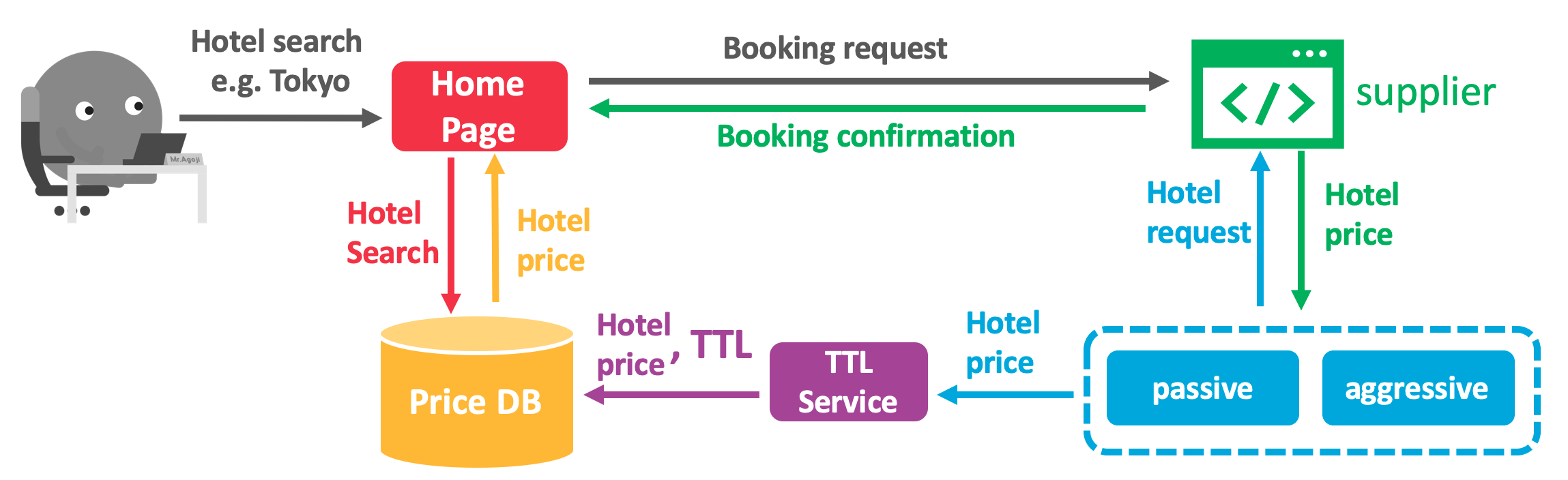}
    \caption{$\superAgg$ system flow. }
    \label{fig:superAgg_flow}
\end{figure}

To overcome the 4 challenges mentioned above, 
we propose $\superAgg$, an intelligent system for hotel price fetching.
As presented in Figure~\ref{fig:superAgg_flow}, 
before every price is written to cache (Price DB), 
it always goes through a TTL service, 
which assigns different TTL for the different hotel searches. 
This TTL service is built on historical data extracted to 
optimize the trade-off between cache hit and price accuracy, 
which addresses the Challenge 1.

Apart from passively sending requests to supplier to fetch for hotel price, 
$\superAgg$ re-invent the process by 
adding an aggressive service which pro-actively 
sends requests to supplier to fetch for hotel price 
on a constant QPS.
By having a constant QPS, 
Challenge 2 and Challenge 3 can be addressed easily. 
Moreover, this aggressive service does not wait for a hotel search 
to appear before sending requests to supplier. 
Therefore, it can increase the cache hit and hence, addresses Challenge 4.

In summary, we make the following contributions in the paper:
\begin{enumerate}
    \item We propose $\superAgg$, 
    an intelligent system which maximizes the bookings for a limited QPS.
    To the best of our knowledge, 
    this is the first productionised intelligent system which optimises the utilization of QPS.
    \item We present a TTL service, 
    SmartTTL which optimizes the trade-off between cache hit and price accuracy.
    \item Extensive A/B experiments were conducted to show that $\superAgg$ is effective and increases \agoda's revenue significantly.
\end{enumerate}

The rest of the paper is organized as following.
Section~\ref{sec:definition} presents the necessary definitions before presenting the TTL service, \smartTTL\ in Section 3. 
In Section 4, we present the aggressive model.
In Section 5, we present the experiment results and analysis. 
Section 6 presents the related work
before concluding the paper in Section 7.

\section{Preliminary and Definition}
\label{sec:definition}
\begin{figure}[t]
    \centering
    \includegraphics[height = 0.65in]{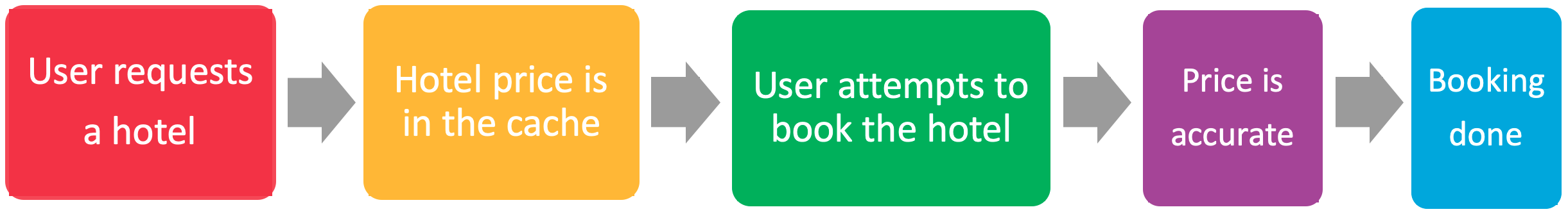}
    \caption{\agoda\ booking flow}
\label{fig:bookinig_flow}
\end{figure}

In this section, we make necessary definitions.
Figure~\ref{fig:bookinig_flow} presents the major steps in the hotel booking process.
In stage 1, an user requests for a hotel price.
In stage 2, if the hotel price is already existing in the cache, then the user will be presented with the cached price. Otherwise, the user won't be able to see the hotel price.
In stage 3, if the user is happy with the hotel price, then the user clicks booking. 
In stage 4, \agoda\ confirms with the hotel whether the price is eligible to sell. If the price is eligible to sell, then \agoda\ confirms the booking in stage 5.

\begin{definition}
Let $U=\{u_1, u_2, \dots, u_{|U|}\}$ 
be the set of users requesting for hotels in \agoda.
Let $H=\{h_1, h_2, \dots, h_{|H|}\}$ 
be the set of hotels that \agoda\ have.
Let $C=\{c_1,c_2,\dots, c_{|C|}\}$ 
be the set of search criteria that \agoda\ receives, 
and each $c_i$ is in the form of 
$ \langle \texttt{checkin,checkout,adults,children,}$
$\texttt{rooms} \rangle$.
\end{definition}

In the definition above, $U$ and $H$ are self-explanatory.
For $C$, $\langle \texttt{2020-05-01,}$
$\texttt{2020-05-02, 2,0,1} \rangle$ 
means a search criteria having the 
checkin date as $\texttt{2020-05}$
$\texttt{-01}$, 
the checkout date as $\texttt{2020-05-02}$,  
the number of adults as $\texttt{2}$, 
the number of children as $\texttt{0}$
and the number of room as $\texttt{1}$.
Therefore, we can define the itinerary request and the user search as follows.

\begin{definition}
Let $R = \{r_1, r_2, \dots, r_{|S|}\}$ be the set of itinerary request that \agoda\ sends to the suppliers, where $r_i \in  H \times C$.
Let $S = \{s_1, s_2, \dots, s_{|S|}\}$ be the set of searches that \agoda\ receives from the user, where $s_i \in U \times H \times C$.
\end{definition}

For example, an itinerary request $r_i = \langle \texttt{Hilton Amsterdam}$,$\texttt{2020-06-01}$, $\texttt{2020-06-02,1,0,1} \rangle$ means \agoda\ sends a request to the supplier to 
fetch price for hotel $\texttt{ Hilton Amsterdam}$ on  $\texttt{checkin=2020-06-01, checkout=}$,\linebreak
$\texttt{2020-06-02}$, 
$\texttt{adults=1}$, 
$\texttt{children=0}$,
$\texttt{rooms=1}$. 

Similarly, an user search 
$s_i= \langle \texttt{Alex}$, 
$\texttt{Hilton Amsterdam}$,
$\texttt{2020-05-01}$,
$\texttt{2020-}$
$\texttt{05-02}$,
$\texttt{2}$,$\texttt{0,1} \rangle$ 
means $\texttt{Alex}$ searched on hotel $\texttt{Hilton Amsterdam}$ for price 
on $\texttt{checkin=2020-05-01}$, 
$\texttt{checkout=2020-05-02}$, 
$\texttt{adults=2}$, 
$\texttt{children=0}$, 
$\texttt{rooms=1}$.  
Note that, if $\texttt{Alex}$ makes the same searches on 
$\texttt{Hilton Amsterdam}$,
$\texttt{2020-05-01}$,
$\texttt{2020-05-02}$,
$\texttt{2}$,$\texttt{0,1}$
multiple times in a day, it is considered as multiple user searches. Therefore, S here is a multi-set.

\begin{definition}
$P_D(s_i)$ is the probability of an user search $s_i$ that hits on the hotel prices in the cache. 
\end{definition}
For example, if $\texttt{Alex}$ makes the 10 searches on 
$\texttt{Hilton Amsterdam}$,
$\texttt{2020-05-01}$,
$\texttt{2020-05-02}$,
$\texttt{2}$,
$\texttt{0,1}$, and 8 out of these 10 searches hit on the price cached. 
Then, $P_D(\langle \texttt{Alex}, 
\texttt{Hilton Amsterdam},
\texttt{2020-05-01},
\texttt{2020-05-02},
\texttt{2},
\texttt{0,1} \rangle ) 
= \frac{8}{10} = 0.8$

\begin{definition}
$P_B(s_i)$ is the probability of an user search $s_i$ that ended up with booking attempt, given that the hotel price is in the cache.
\end{definition}

Following the above example, for $\langle \texttt{Alex}$, 
$\texttt{Hilton Amsterdam}$,
$\texttt{2020-05-01}$,
$\texttt{2020-05-02},
\texttt{2},
\texttt{0,1} \rangle$, 
 $\texttt{Alex}$ has 8 searches returned prices. 
 And out of these 8 searches, 
 $\texttt{Alex}$ makes 2 booking attempts.
 Then, $P_B(\langle \texttt{Alex}, 
\texttt{Hilton Amsterdam}$,
$\texttt{2020-05-01},
\texttt{2020-05-02},
\texttt{2},
\texttt{0,1} \rangle) = \frac{2}{8} = 0.25$

\begin{definition}
$P_A(s_i)$ is the probability of the hotel price is accurate after an user makes a booking attempt on search $s_i$. 
\end{definition}

Continuing the example above, 
out of the 2 booking attempts, 
1 booking attempt succeeds. 
Hence, $P_A(\langle \texttt{Alex}, 
\texttt{Hilton Amsterdam}$,
$\texttt{2020-05-01},
\texttt{2020-05-02}$, 
$\texttt{2},\texttt{0,1} \rangle)$ = $\frac{1}{2} = 0.5$.
Therefore, we can formulate the number of bookings expected as follows.

\begin{definition}
The expected number of bookings is the following
\begin{equation}
    K = \sum_{s_i} P_D(s_i) \times P_B(s_i) \times P_A(s_i)
    \label{eqn:booking}
    \end{equation}
\end{definition}

Therefore, our goal is to optimise such $K$.
To optimize $K$, 
we would expect $P_D(s_i)$, $P_B(s_i)$, $P_A(s_i)$ to be as high as possible.
$P_B(s_i)$ is an user behaviour, 
as a hotel price fetching system, 
this is not controllable. 
But we can learn this $P_B$ from historical data.
However, $P_D(s_i)$, 
$P_A(s_i)$ could be tuned by adjusting the TTL. 
As illustrated by Figure\ref{fig:ttl_role},
to increase $P_D(s_i)$, one can simply increase the TTL.
Similarly, to increase $P_A(s_i)$, 
one just needs to decrease TTL. 
We will discuss how to set the TTL to optimize the booking in Section~\ref{sec:smartTTL}.


\section{SmartTTL}
\label{sec:smartTTL}
In this section,
we explain how we build a smart TTL service which 
assigns itinerary request specific TTL to optimize the bookings. 
There are three major steps:  price-duration extraction, 
price-duration clustering and TTL assignment.

\subsection{Price-Duration Extraction}
Price-duration refers to how long each price stay unchanged. This is approximated by the time difference between two consecutive requests of the same itinerary that \agoda\ sends to the supplier. Figure 7 presents an example of extracting price-duration distribution from empirical data of hotel $\texttt{Hilton}$
$\texttt{Amsterdam}$ and search criteria 
$\langle \texttt{2019-10-01,2019-10-02}$, $\texttt{1,0,1} \rangle$.

\agoda\ first sends a request to supplier at $\texttt{13:00}$ to fetch for price, 
and that's the first time we fetch price for such itinerary.
So, there is no price change and no price-duration extracted.
Later, at $\texttt{13:31}$, 
\agoda\ sends the second request to the supplier to fetch for price, 
and observes that the price has changed. 
Hence, the price-duration for the previous price is 31 minutes 
(the time difference between \texttt{13:00} and \texttt{13:31}).
Similarly, at $\texttt{14:03}$, 
\agoda\ sends the third request to the supplier to fetch for price, 
and again, observes that the price has changed. 
Hence, the price-duration for the second price is 32 minutes.
Therefore, for each search criteria, we can extract the empirical price-duration distributions.

\begin{figure}
      \centering
    \includegraphics[height = 1.8in]{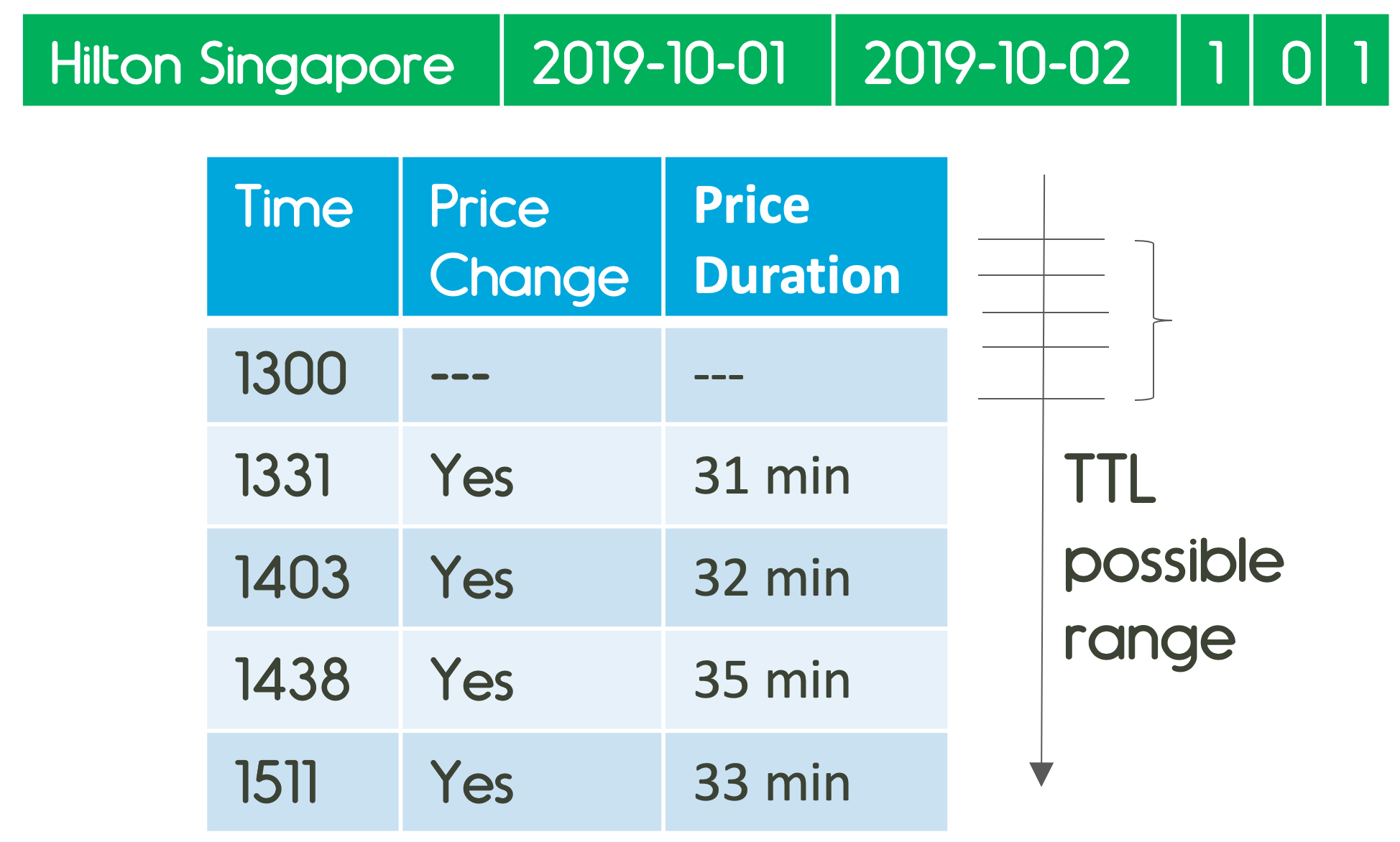}
    \caption{Price duration extraction from empirical data}
    \label{fig:ttl_extract}
   \end{figure}

\subsection{Price-Duration Clustering}
In \agoda, we have billions of such user searches every day. 
It is practically intractable and unnecessary 
to store such volume of search criteria's TTL into in-memory cache, e.g. Redis or Memcached.
Therefore, we need to reduce the cardinality of the user searches. 
And we do it through clustering.

\begin{figure}[h]
    \centering
    \includegraphics[height = 1.4in]{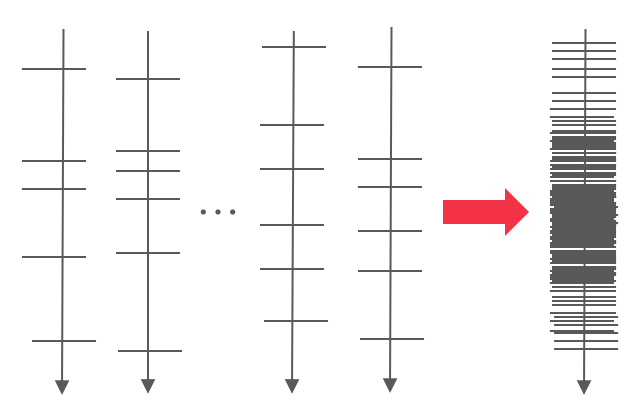}
    \caption{Similar supplier requests are clustered together}
    \label{fig:ttl_cluster}
\end{figure}

Figure~\ref{fig:ttl_cluster} presents the price-duration clustering process.
We cluster these user searches into clusters to reduce the cardinality. 
In $\superAgg$, we used $\xgboost$~\cite{xgboost}
for the clustering feature ranking, 
and the significant features are $\texttt{checkin}$, $\texttt{price\_availability}$. 
We observe that the itinerary requests with same $\texttt{checkin}$ and $\texttt{price\_availability}$ (whether the hotel is sold out or not) 
have the similar price-duration distribution.
Hence, for all supplier requests with same $\texttt{checkin}$ and $\texttt{price\_availability}$, we group them into the same cluster, and use the aggregated price-duration distribution to represent the cluster.
By doing this, we dramatically reduce the cardinality to $\sim1000$, 
which can be easily stored into any in-memory data structure.

\subsection{TTL Assignment}
In the above section, 
we finished clustering.
Next, we need to assign a TTL for each cluster.
Note that, we want to optimize the bookings as expressed in Equation 1, 
and the TTL will affect the cache hit ($P_D$ in Equation 1) 
and booking price ($P_A$ in Equation 1) accuracy. 
Hence, we want to assign a TTL for each cluster in which Equation 1 is optimised.

\begin{figure}[h]
    \centering
    \includegraphics[height=2in]{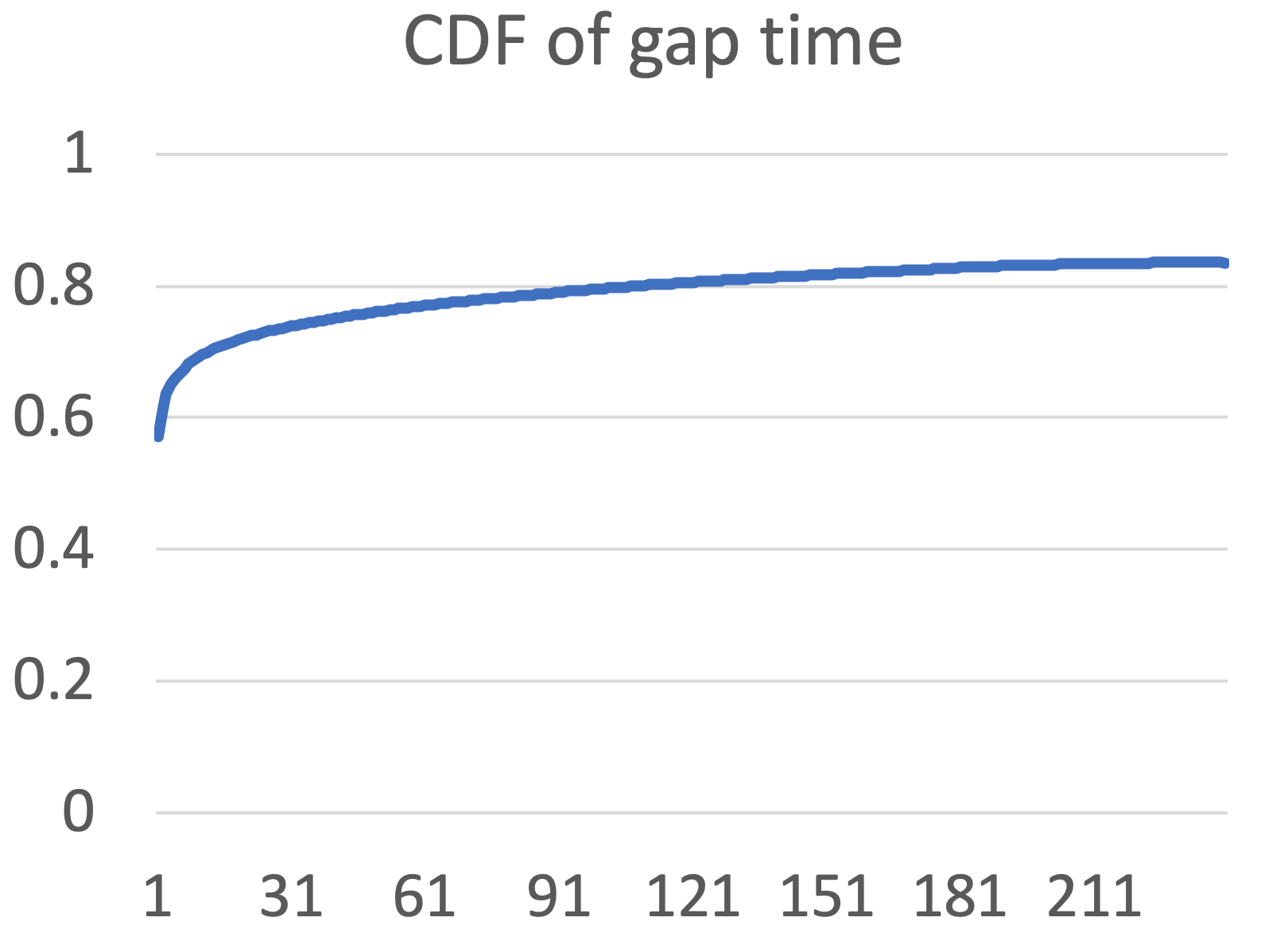}
    \caption{CDF of gap time. x-axis is the gap time in minutes.}
    \label{fig:cdf}
\end{figure}

For cache hit, we can easily approximate the cache miss ratio curve ~\cite{WWW2020} using Cumulative Distribution Function (CDF) of the gap time (time difference between current request and previous request for the same itinerary search).
Figure~\ref{fig:cdf} presents the CDF 
of the gap time, where the x-axis is the gap time, 
and the y-axis is the portion of requests whose gap time $\leq$ a specific gap time. 
For example, $80\%$ of the requests are having gap time $\leq$ 120 minutes in Figure~\ref{fig:cdf}.
Hence, by setting TTL at $120$ minutes, we can achieve $80\%$ cache hit. 
Therefore, the cache miss ratio curve related to TTL can be easily found, 
and we can know the approximated cache hit rate for each TTL we choose.

For booking accuracy of a cluster $C$, this can be approximated by 
$$
\frac{\sum_{r_i \in C } \min(1,\frac{TTL_{r_i}}{TTL_{assigned}} )}{|C|}
$$
For example, in a specific cluster, 
if the empirical price-duration observed is $120$ minutes and $100$ minutes, and we assigned $150$ minutes. 
Hence, we know that there are $120$ and $100$ minutes that we are using the accurate price.
Hence, the accuracy is $(\frac{120}{150} + \frac{100}{150})/2 = \frac{11}{15}$.

Hence, to optimize the bookings as expressed in Equation~\ref{eqn:booking}, 
we just need to numerate the different TTL in each cluster to find such TTL.

So far, we have completed the major steps in SmartTTL.

\section{From Passive Model to Aggressive Model}
As mentioned in Section 1, \smartTTL addresses the Challenge 1. 
We still have three more challenges remaining untackled.
For Challenge 2 and Challenge 3, 
we can resolve it by guaranteeing each data centre sends constant rate $\mu$ of requests to the suppliers. 
Every time passive model sends $\mu_{passive}$ requests to the suppliers, 
where $\mu_{passive} \textless \mu$, 
we proactively send extra $\mu - \mu_{passive}$ requests to supplier.
The question is how to generate these $\mu - \mu_{passive}$ requests.
Next, 
we will present one alternative of generating such $\mu - \mu_{passive}$ requests.

\subsection{Aggressive Model with LRU Cache}
In this section, 
we describe an aggressive model which aggressively sends requests to the supplier to fetch for hotel price. 
These requests are generated from the auxiliary cache $\C_{LRU}$. There are two major steps:

\textbf{Cache building}. 
    The auxiliary cache $\C_{LRU}$ is built up by using historical user searches. 
    For each user search $s_i$, they are always admitted into $\C_{LRU}$.
    Once $\C_{LRU}$ reaches its maximum capacity specified, 
    $\C_{LRU}$ will evict the user search $s_i$ which is Least Recently Used (LRU).
    
\textbf{Request pulling}.
    At every second $t_i$, 
    passive model needs to send $\mu_{passive}$ requests to supplier. 
    And the supplier allows us to send $\mu$ requests per second. 
    Hence, 
    aggressive model will send $\mu_{aggressive} = \mu - \mu_{passive}$ requests to the supplier.
    To generate such $\mu_{aggressive}$ requests, 
    \agoda\ pulls $\mu_{aggressive}$ requests from $\C_{LRU}$ which are going to expire (starts from requests that are closets to expiry until $\mu_{aggressive}$ is used up).

It is obvious that the above approach can solve Challenge 2 and Challenge 3. 
Moreover, it can also help improve the cache hit by requesting the hotel prices 
before an user searches for it. 

However, this is not optimal. 
For example, 
a specific hotel could be very popular. 
However, if the hotel is not price competitive, 
then \agoda\ does not need to waste such QPS to 
pull the hotel price from such supplier.
In the next section, we will introduce an aggressive model which optimizes the bookings.

\subsection{Aggressive model with SmartScheduler}
As mentioned, aggressive model with LRU cache is not optimal.
Moreover, previously, passive model always has the highest priority.
Meaning aggressive model only sends requests to supplier if there is extra QPS left.
However, this is again not optimized.
In this section, we present an aggressive model which optimizes the bookings.
It has 5 major steps.

\textbf{Itinerary frequency calculation}. This describes how many times an itinerary needs to be requested to ensure it is always available in database.
If we want a high cache hit rate, we want an itinerary $r_i$ to be always available in the database,
that means we need to make sure that such itinerary $r_i$ is fetched before it expires.
Moreover, for each $r_i$, 
we have the generated $TTL_{r_{i}}$.
Hence, to make sure an itinerary $r_i$ is always available in database $D$ 
for 24 hours (1440 minutes), 
we need to send $f_{r_i}$ requests to supplier, where $f_{r_i}$ is 
\begin{equation}
f_{r_i}=\left \lceil \frac{1440}{TTL_{r_i}} \right \rceil
\label{eqn:itinerary_frequency}
\end{equation}

\textbf{Itinerary value evaluation}.This evaluates the value of an itinerary by the probability of booking from this itinerary.
With above itinerary frequency calculation, we can assume an itinerary request is always a 'hit' in the database. Hence, in this step, we evaluate the itinerary value given that such itinerary is always available in our Price DB.
That is, for all user search $s_i$ on the same itinerary $r_i$, $s_i \succ r_i$, 
it will be always cache hit, i.e. $\pd=1$.
Recall from Equation~\ref{eqn:booking},
for each itinerary request $r_i$, 
we have now the expected number of bookings as
\begin{equation}
    E[K_{r_i}] = \sum_{s_i \in r_i} \pd \times \pb \times \pa = \sum_{s_i \in r_i} \pb \times \pa
    \label{eqn:itinerary_value}
\end{equation}

\textbf{Request value evaluation}. This evaluates the value of a request by the probability of booking from this request.
By Equation~\ref{eqn:itinerary_value} and Equation~\ref{eqn:itinerary_frequency},
we can have the expected bookings per supplier request as 
\begin{equation}
\frac{E[K_{r_i}]}{f_{r_i}}
\label{eqn:bookingperrequest}
\end{equation}

\textbf{Top request generation}. This generates the top requests we want to select according to their values.
Within a day, for a specific supplier, 
we are allowed to send $M=\mu \times 60 \times 60 \times 24$ requests to supplier.
Therefore, by Equation~\ref{eqn:bookingperrequest}, 
we can order the supplier requests and pick the most valuable $M$ requests.

\textbf{Top request scheduling}. This describes how to schedule to pull the top requests we selected.
Given that we have $M$ requests need to be sent to the supplier,
we need to make sure 1. each of these requests is sent to the supplier 
before its previous request expires.
    2. at every second, we send exactly $\mu$ requests to the supplier.
    
For all itinerary requests, we group the itinerary request by their frequency,
where $G(f_i) = [r_1, r_2, r_3, \dots, r_k]$ and itinerary request $r_1, r_2, r_3, \dots, r_k$ all have frequency $f_i$ and same $TTL_{i}$. This means every single request $r_i$ where $i = 1, 2, 3, ..., k$, is to be scheduled to send for $f_i$ times and all k itinerary requests, $r_1, r_2, r_3, \dots, r_k$, are to be sent within a period of $TTL_{i}$.
To ensure every of these k itinerary requests are sent within a period of $TTL_{i}$, we can simply distribute these $r_1, r_2, r_3, \dots, r_k$ requests evenly over each second in $TTL_{i}$. Thus, we need to schedule to send $\frac{k}{TTL_i}$ requests each second within $TTL_i$. Now we just need to send the same set of requests every $TTL_i$ and repeats this process for $f_i$ times For example, if $G(4)=[r_1, r_2, \dots, r_{43200}]$, 
then we have 43200 itinerary requests having frequency = $4$ and $TTL$ = $6$ hours which is 21600 seconds. That means, in every 6 hours, we need to schedule 43200 itinerary requests, which is $\frac{43200}{21600} = 2$ requests per second. That is, if we don't consider any ordering of the 43200 requests, we will send requests $r_1$ and $r_2$ in 1st second, $r3$ and $r4$ in 2nd second until $r_{43199}$ and $sr_{43200}$ itinerary requests in 21600th seconds. In 21601th second, $r_1$ and $r_2$ will be sent again and so on. These 43200 itinerary requests are scheduled to be sent for a frequency of 4 times in a single day.

By having the above 5 steps, we can see that the most valuable $M$ requests are sent by SmartScheduler which maximizes the booking.

 





\section{Experiment and Analysis}
The aggressive model with SmartScheduler has been deployed in production at \agoda. 
The deployment has yielded significant gains on bookings and other system metrics. 
Before the deployment, 
we have done extensive online A/B experiment to evaluate the effectiveness of the model. 

In the following section, we will present the experiments conducted in 2019. 
As \agoda\ is a publicly listed company, 
we are sorry that we can't reveal the exact number of bookings due to data sensitivity, 
but we will try to be as informative as possible.
Overall, aggressive model with SmartScheduler wins other baseline algorithms by $10\%$ to $30\%$.

\subsection{Experimentation suppliers}
There are two types of suppliers \agoda\ have experimented with:
\begin{enumerate}
    \item \textbf{Retailers}. 
    Retailer are those suppliers whose market manager from each OTA deals with hotel directly and they are selling hotel rooms online.
    \item \textbf{Wholesalers}. Wholesalers are those suppliers that sell hotels/hotel’s room/other products in large quantities at lower price (package rate), mainly selling to B2B or retailer not direct consumer.
\end{enumerate}

In this paper, we present the results from 5 suppliers.

\textit{ \textbf{Supplier A} } is a Wholesaler supplier which operates in Europe.

\textit{ \textbf{Supplier B} } is a Wholesaler supplier which operates worldwide.

\textit{ \textbf{Supplier C} } is a Wholesaler supplier which operates worldwide.

\textit{ \textbf{Supplier D} } is a Retailer supplier which operates in Japan.

\textit{ \textbf{Supplier E} } is a Retailer supplier which operates in Korea.

In this section, all the experiments were conducted through online A/B experiment over 14 days, 
where half of the allocated users are experiencing algorithm A and 
the other half are experiencing algorithm B.
Moreover, for all the plots in this section,
\begin{itemize}
    \item x-axis is the nth day of the experiment.
    \item bar-plot represents the bookings and line-plot represents the cache hit.
\end{itemize}

\subsection{Fixed TTL v.s. SmartTTL}
In this section, we compare the performance between passive model with Fixed TTL (A) and passive model with SmartTTL (B).
\begin{figure}
    \centering
    \includegraphics[height=1.8in]{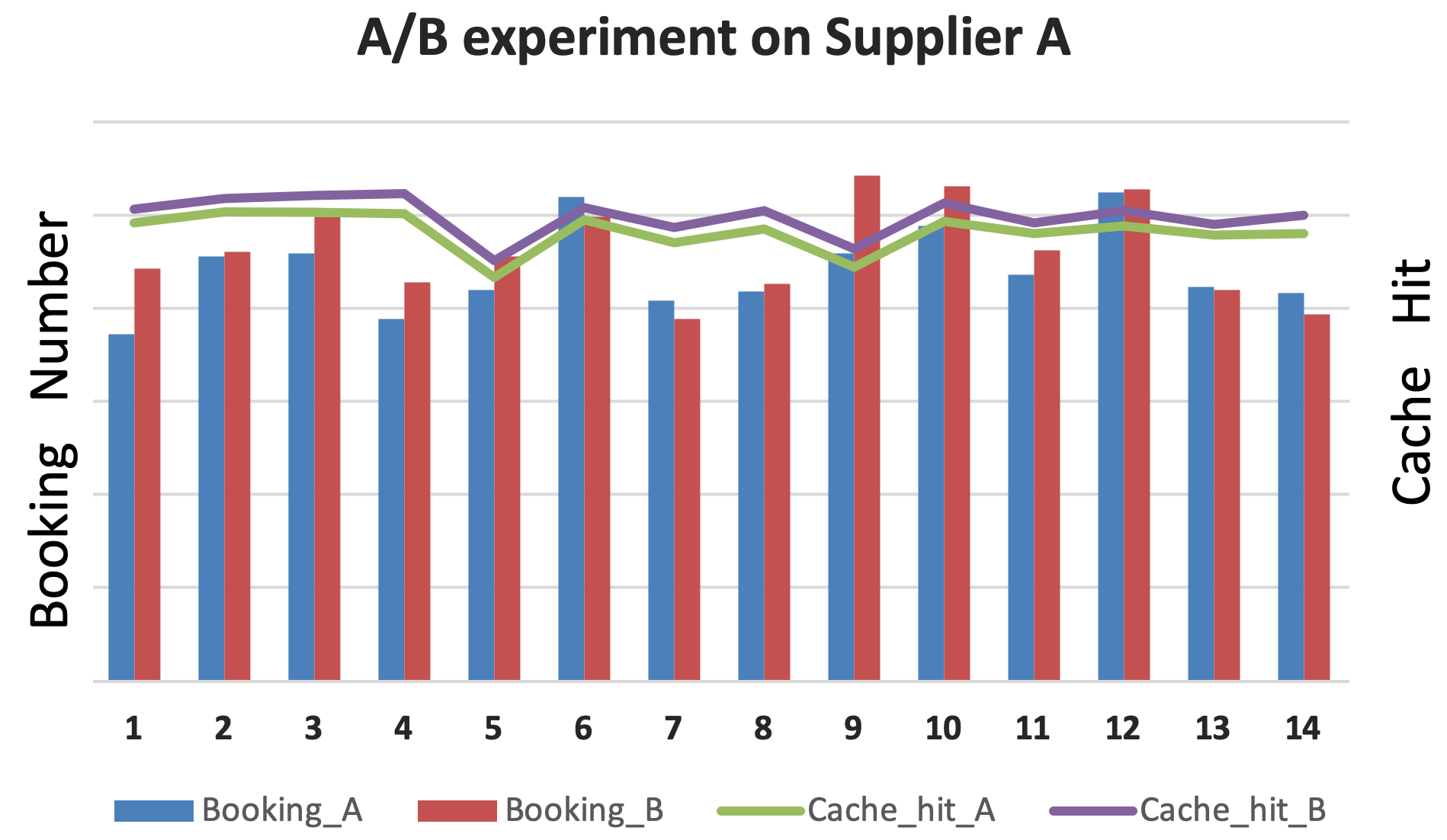}
    \caption{A/B Experiment on Supplier A}
    \label{fig:expa}
\end{figure}
Figure~\ref{fig:expa} presents the results on Supplier A, 
and we can see that B variant wins A variant by a small margin. 
Overall, B variant wins by $2-4\%$ for cache hit, and $\sim2\%$ for bookings. This is expected as SmartTTL only address Challenge 1.

\subsection{SmartTTL v.s. Aggressive Model with SmartScheduler}
In this section, we compare the performance between passive model with SmartTTL (A) and aggreesive model with SmartScheduler (B).
We present the A/B experiment results
Supplier C and Supplier E.

\begin{figure}
    \centering
    \includegraphics[height=1.8in]{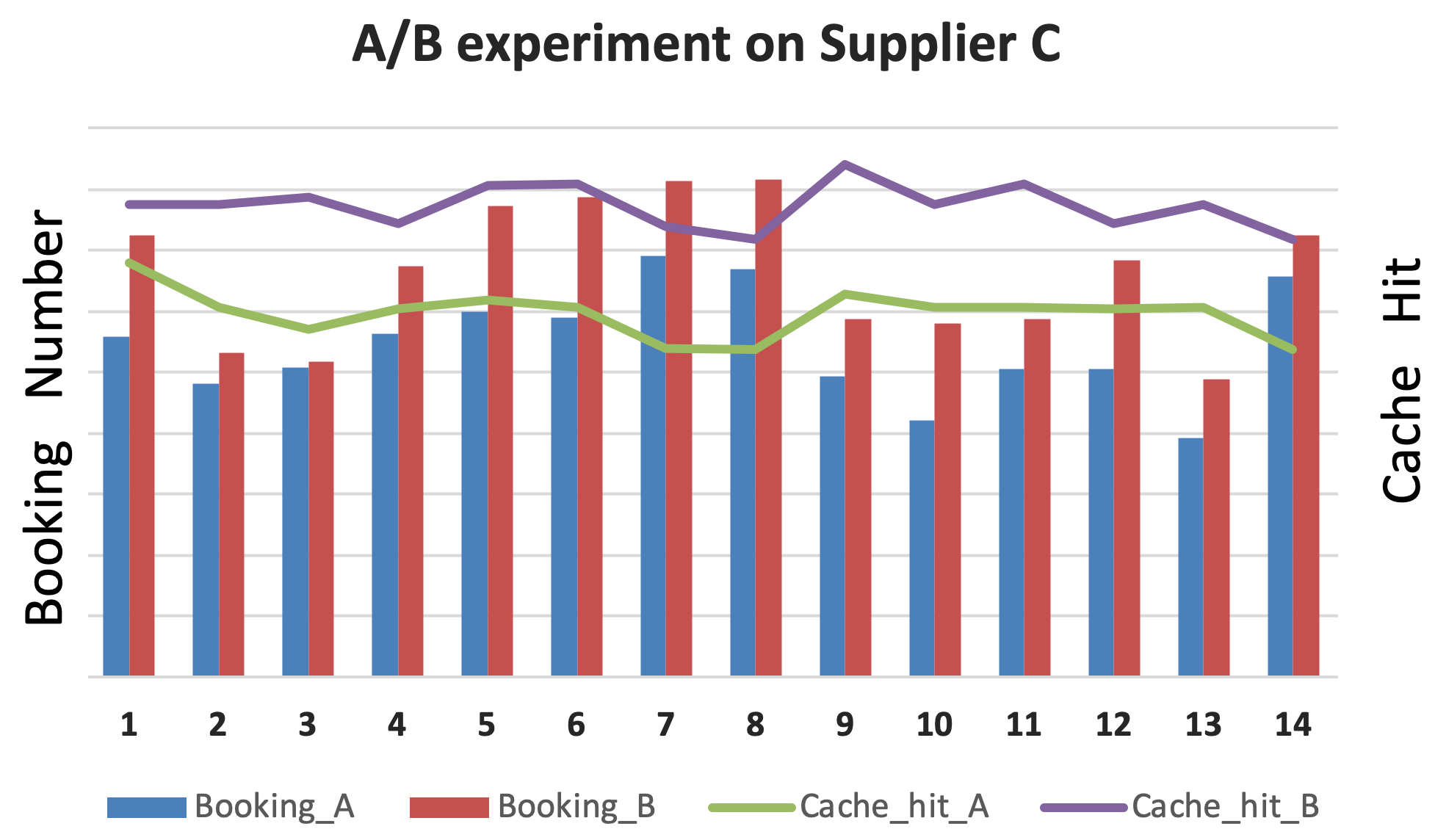}
    \caption{A/B Experiment on Supplier C}
    \label{fig:expc}
\end{figure}

Figure~\ref{fig:expc} presents the results on Supplier C, 
and we can easily see that B variant wins A variant significantly in terms of booking and cache hit ratio. 
For cache hit and bookings, B variant wins A variant consistently.

\begin{figure}
    \centering
    \includegraphics[height=1.8in]{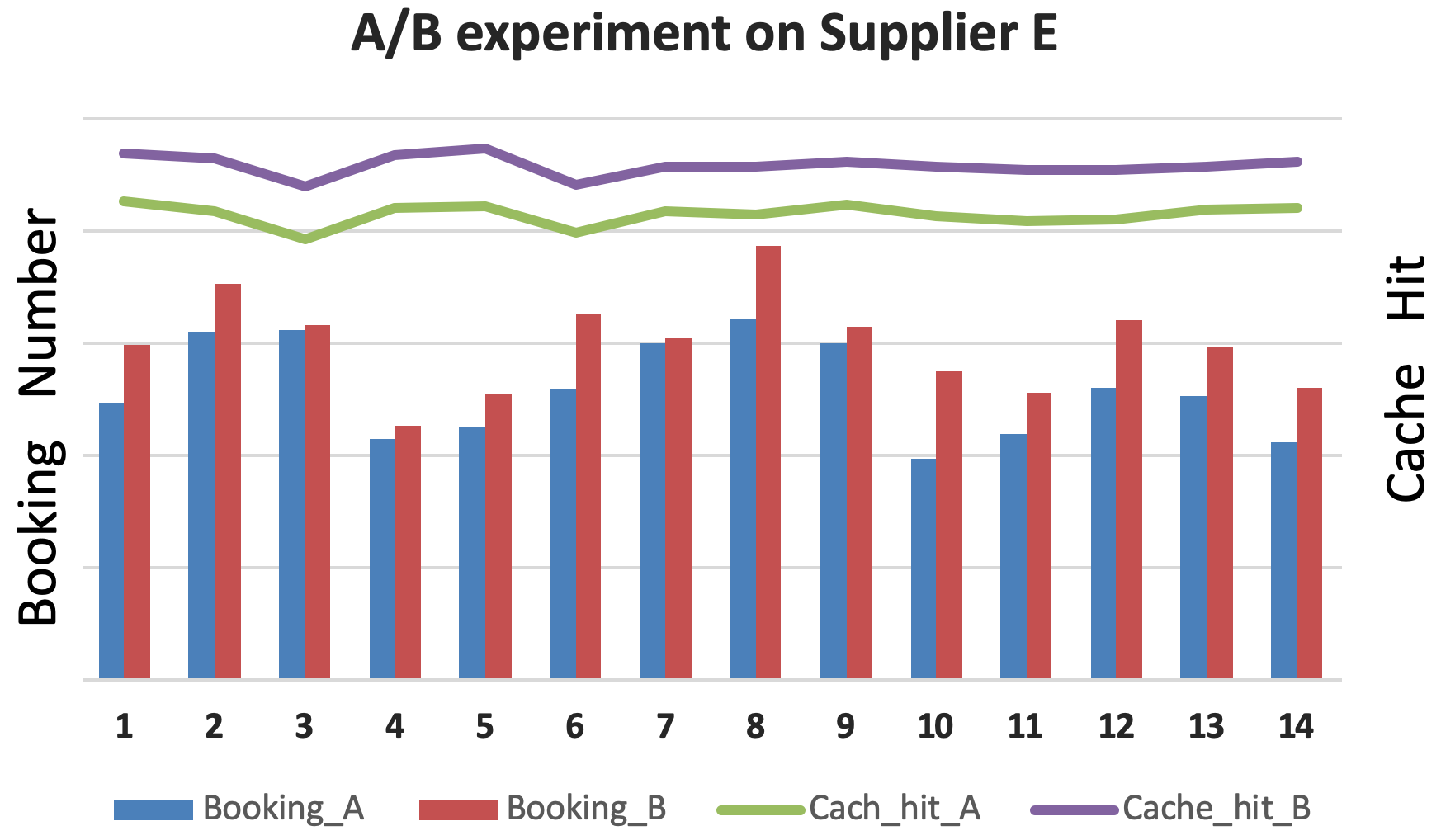}
    \caption{A/B Experiment on Supplier E }
    \label{fig:expe}
\end{figure}
Figure~\ref{fig:expe} presents the results on Supplier E, 
and we can easily see that B variant wins A variant significantly in terms of booking and cache hit ratio. 
For cache hit, B variant wins A variant consistently.
For bookings, we can see that B never lose to A on any single day.

\subsection{Aggressive Model with LRU Cache v.s. Aggressive Model with SmartScheduler}
In this section, we compare the performance between aggressive model with LRU cache (A) and aggreesive model with SmartScheduler (B).
We present the A/B experiment results
Supplier B and Supplier D. 

\begin{figure}
    \centering
    \includegraphics[height=1.8in]{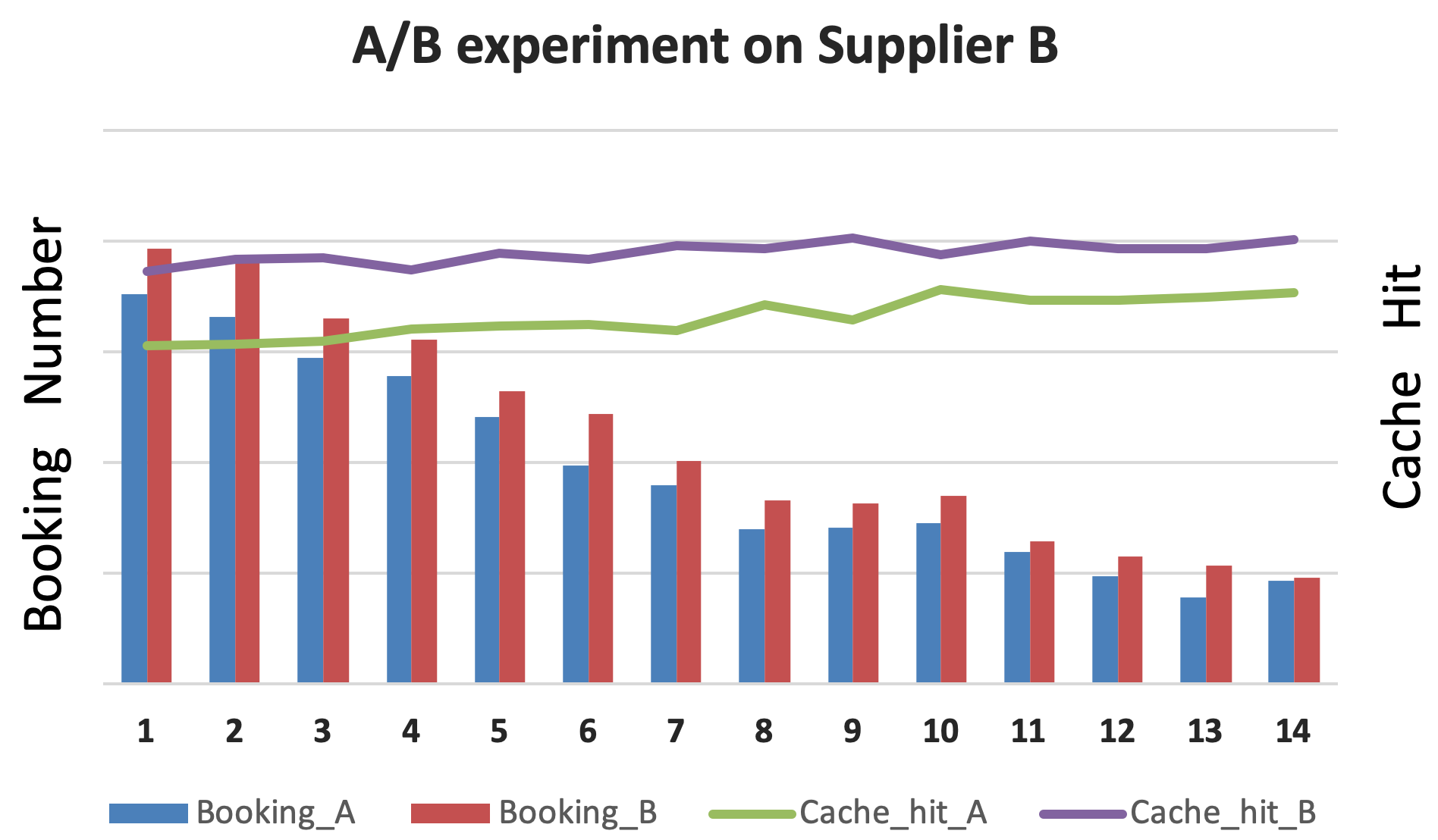}
    \caption{A/B Experiment on Supplier B}
    \label{fig:expb}
\end{figure}

Figure~\ref{fig:expb} presents the results on Supplier B, 
and we can easily see that B variant wins A variant significantly in terms of booking and cache hit ratio. 
For cache hit, B variant wins A variant consistently.
It is worthwhile to note that the overall booking declines along the x-axis, 
this could be caused by many factors such as promotions from competitors, seasonality etc. 
However, B variant is still able to win A variant by a consistent trend.

\begin{figure}
    \centering
    \includegraphics[height=1.8in]{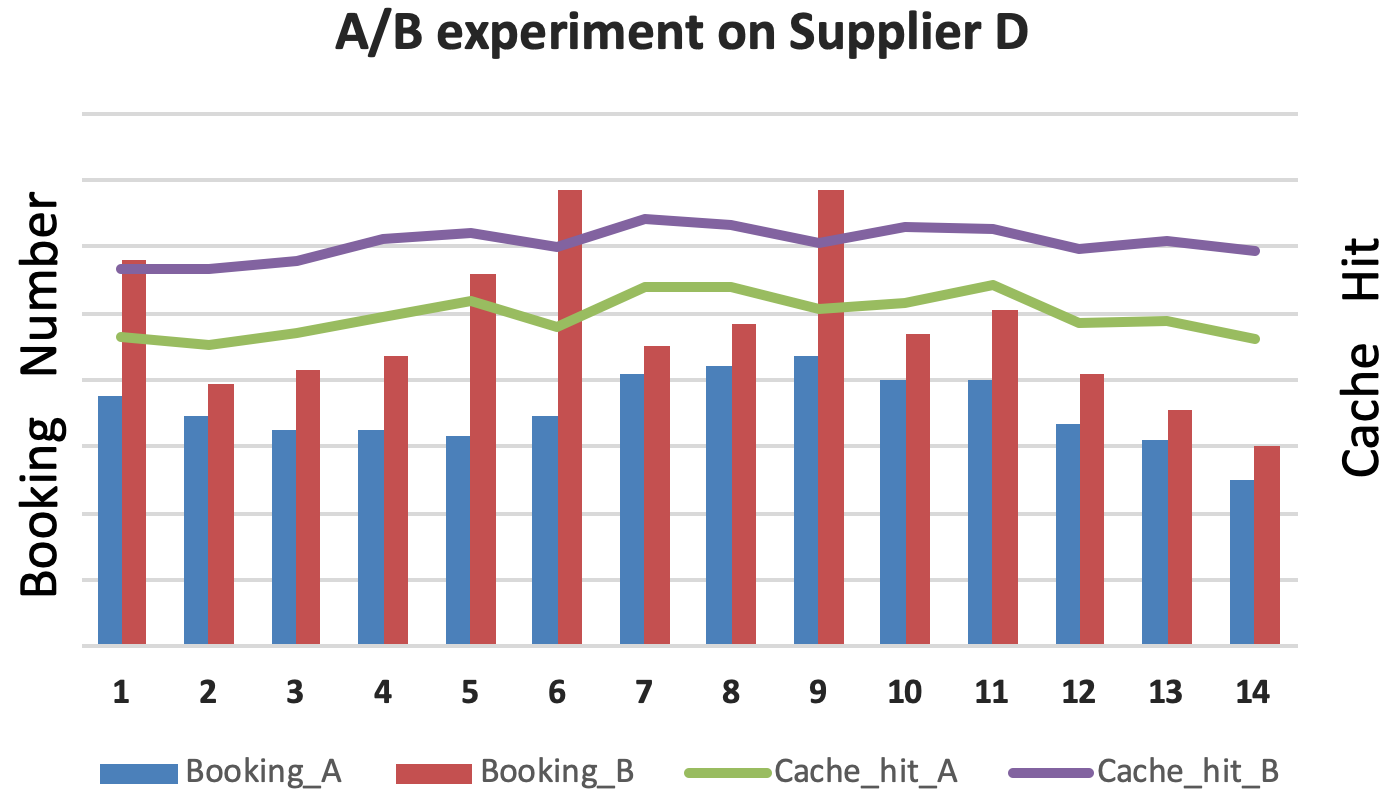}
    \caption{A/B Experiment on Supplier D}
    \label{fig:expd}
\end{figure}

Figure~\ref{fig:expd} presents the results on Supplier D, 
and we can easily see that B variant wins A variant significantly in terms of booking and cache hit ratio. 
For cache hit, B variant wins A variant consistently.
For bookings, we can see that B consistently wins A by more than $10\%$. 
And on certain days, e.g. day 5, B wins by more than $50\%$.

\section{Related Work}
The growth of traveling industry has attracted substantial academic attention~\cite{airbnb,bcom, europricing}. 
To increase the revenue, many effort have been spent on enhancing the pricing strategy. 

Aziz et al. proposed a revenue management system framework based on price decisions which optimizes the revenue~\cite{roompricing}.
Authors in ~\cite{airbnb} proposed Smart Price which improves the room booking by guiding the hosts to price the rooms in Airbnb.
As long-term stay is getting more common, Ling et al. ~\cite{long_stay} derived the optimal pricing strategy for long-term stay, which is beneficial to hotel as well as its customer. 
Similar efforts have been seen in ~\cite{noone2016pricing,dynamicpricing} in using pricing strategies to increase the revenues.

Apart from pricing strategy, 
some effort has been spent on overbooking~\cite{toh2002hotel,koide}. 
For example, Antonio et al.~\cite{ANTONIO2017} built prediction models for predicting cancellation of booking to mitigate revenue loss derived from booking cancellations.

Nevertheless, none of the existing work has studied hotel price fetching strategy. 
To the best of our knowledge, we are the first to deploy an optimized price fetching strategy which increases the revenue by large margin.

\section{Conclusion and Future Work}
In this paper, we presented $\superAgg$, 
an intelligent hotel price fetching system which optimizes the bookings.
To the best of our knowledge,\\ 
$\superAgg$ is the first productionized system which addresses the 4 challenges mentioned in Section 1.
It differs from most existing OTA system by having SmartTTL which determines itinerary specific TTL.
Moreover, instead of passively sending requests to suppliers, 
$\superAgg$ aggressively fetches the most valuable hotel prices from suppliers which optimizes the bookings.
Extensive online experiments shows that 
$\superAgg$ is not only effective in improving system metrics like cache hit, 
but also grows the company revenues significantly. 
We believe that $\superAgg$ is a rewarding direction for application of data science in OTAs.

One of the factor which brings bookings is pricing.
In the future, we will explore how to optimize the bookings through a hybrid of pricing strategy and pricing fetching strategy.

\tiny
\printbibliography

\end{document}